\documentclass[prc,twocolumn,showpacs,amsmath,amssymb,floatfix,superscriptaddress]{revtex4-1}

\usepackage{lineno}
\usepackage{graphicx}
\usepackage{dcolumn}
\usepackage{bm}

\begin{document}

\preprint{Version 2.23}

\title{Balance Functions from Au+Au, d+Au, and p+p Collisions at $\sqrt{s_{NN}}$ = 200 GeV}

\affiliation{Argonne National Laboratory, Argonne, Illinois 60439, USA}
\affiliation{University of Birmingham, Birmingham, United Kingdom}
\affiliation{Brookhaven National Laboratory, Upton, New York 11973, USA}
\affiliation{University of California, Berkeley, California 94720, USA}
\affiliation{University of California, Davis, California 95616, USA}
\affiliation{University of California, Los Angeles, California 90095, USA}
\affiliation{Universidade Estadual de Campinas, Sao Paulo, Brazil}
\affiliation{University of Illinois at Chicago, Chicago, Illinois 60607, USA}
\affiliation{Creighton University, Omaha, Nebraska 68178, USA}
\affiliation{Czech Technical University in Prague, FNSPE, Prague, 115 19, Czech Republic}
\affiliation{Nuclear Physics Institute AS CR, 250 68 \v{R}e\v{z}/Prague, Czech Republic}
\affiliation{University of Frankfurt, Frankfurt, Germany}
\affiliation{Institute of Physics, Bhubaneswar 751005, India}
\affiliation{Indian Institute of Technology, Mumbai, India}
\affiliation{Indiana University, Bloomington, Indiana 47408, USA}
\affiliation{Alikhanov Institute for Theoretical and Experimental Physics, Moscow, Russia}
\affiliation{University of Jammu, Jammu 180001, India}
\affiliation{Joint Institute for Nuclear Research, Dubna, 141 980, Russia}
\affiliation{Kent State University, Kent, Ohio 44242, USA}
\affiliation{University of Kentucky, Lexington, Kentucky, 40506-0055, USA}
\affiliation{Institute of Modern Physics, Lanzhou, China}
\affiliation{Lawrence Berkeley National Laboratory, Berkeley, California 94720, USA}
\affiliation{Massachusetts Institute of Technology, Cambridge, MA 02139-4307, USA}
\affiliation{Max-Planck-Institut f\"ur Physik, Munich, Germany}
\affiliation{Michigan State University, East Lansing, Michigan 48824, USA}
\affiliation{Moscow Engineering Physics Institute, Moscow Russia}
\affiliation{City College of New York, New York City, New York 10031, USA}
\affiliation{NIKHEF and Utrecht University, Amsterdam, The Netherlands}
\affiliation{Ohio State University, Columbus, Ohio 43210, USA}
\affiliation{Old Dominion University, Norfolk, VA, 23529, USA}
\affiliation{Panjab University, Chandigarh 160014, India}
\affiliation{Pennsylvania State University, University Park, Pennsylvania 16802, USA}
\affiliation{Institute of High Energy Physics, Protvino, Russia}
\affiliation{Purdue University, West Lafayette, Indiana 47907, USA}
\affiliation{Pusan National University, Pusan, Republic of Korea}
\affiliation{University of Rajasthan, Jaipur 302004, India}
\affiliation{Rice University, Houston, Texas 77251, USA}
\affiliation{Universidade de Sao Paulo, Sao Paulo, Brazil}
\affiliation{University of Science \& Technology of China, Hefei 230026, China}
\affiliation{Shandong University, Jinan, Shandong 250100, China}
\affiliation{Shanghai Institute of Applied Physics, Shanghai 201800, China}
\affiliation{SUBATECH, Nantes, France}
\affiliation{Texas A\&M University, College Station, Texas 77843, USA}
\affiliation{University of Texas, Austin, Texas 78712, USA}
\affiliation{Tsinghua University, Beijing 100084, China}
\affiliation{United States Naval Academy, Annapolis, MD 21402, USA}
\affiliation{Valparaiso University, Valparaiso, Indiana 46383, USA}
\affiliation{Variable Energy Cyclotron Centre, Kolkata 700064, India}
\affiliation{Warsaw University of Technology, Warsaw, Poland}
\affiliation{University of Washington, Seattle, Washington 98195, USA}
\affiliation{Wayne State University, Detroit, Michigan 48201, USA}
\affiliation{Institute of Particle Physics, CCNU (HZNU), Wuhan 430079, China}
\affiliation{Yale University, New Haven, Connecticut 06520, USA}
\affiliation{University of Zagreb, Zagreb, HR-10002, Croatia}

\author{M.~M.~Aggarwal}\affiliation{Panjab University, Chandigarh 160014, India}
\author{Z.~Ahammed}\affiliation{Lawrence Berkeley National Laboratory, Berkeley, California 94720, USA}
\author{A.~V.~Alakhverdyants}\affiliation{Joint Institute for Nuclear Research, Dubna, 141 980, Russia}
\author{I.~Alekseev~~}\affiliation{Alikhanov Institute for Theoretical and Experimental Physics, Moscow, Russia}
\author{J.~Alford}\affiliation{Kent State University, Kent, Ohio 44242, USA}
\author{B.~D.~Anderson}\affiliation{Kent State University, Kent, Ohio 44242, USA}
\author{D.~Arkhipkin}\affiliation{Brookhaven National Laboratory, Upton, New York 11973, USA}
\author{G.~S.~Averichev}\affiliation{Joint Institute for Nuclear Research, Dubna, 141 980, Russia}
\author{J.~Balewski}\affiliation{Massachusetts Institute of Technology, Cambridge, MA 02139-4307, USA}
\author{L.~S.~Barnby}\affiliation{University of Birmingham, Birmingham, United Kingdom}
% ...

\collaboration{STAR Collaboration}\noaffiliation

\date{\today}

\begin{abstract}
Balance functions have been measured for charged particle pairs, identified charged pion pairs, and identified charged kaon pairs in Au+Au, d+Au, and p+p collisions at $\sqrt{s_{NN}}$ = 200 GeV at the Relativistic Heavy Ion Collider using the STAR detector.  These balance functions are presented in terms of relative pseudorapidity, $\Delta \eta$, relative rapidity, $\Delta y$, relative azimuthal angle, $\Delta \phi$, and invariant relative momentum, $q_{\rm inv}$.  In addition, balance functions are shown in terms of the three components of $q_{\rm inv}$: $q_{\rm long}$, $q_{\rm out}$, and $q_{\rm side}$.  For charged particle pairs, the width of the balance function in terms of $\Delta \eta$ scales smoothly with the number of participating nucleons, while HIJING and UrQMD model calculations show no dependence on centrality or system size.  For charged particle and charged pion pairs, the balance functions widths in terms of $\Delta \eta$ and $\Delta y$ are narrower in central Au+Au collisions than in peripheral collisions. The width for central collisions is consistent with thermal blast-wave models where the balancing charges are highly correlated in coordinate space at breakup. This strong correlation might be explained either by delayed hadronization or by limited diffusion during the reaction. Furthermore, the narrowing trend is consistent with the lower kinetic temperatures inherent to more central collisions. In contrast, the width of the balance function for charged kaon pairs in terms of $\Delta y$ shows little centrality dependence, which may signal a different production mechanism for kaons.  The widths of the balance functions for charged pions and kaons in terms of $q_{\rm inv}$ narrow in central collisions compared to peripheral collisions, which may be driven by the change in the kinetic temperature.
\end{abstract}

\pacs{25.75.Gz}
\maketitle

\section{Introduction}

The study of correlations and fluctuations can provide evidence for the production of a strongly interacting quark-gluon plasma (QGP) in relativistic heavy-ion collisions \cite{stephanov_fluc_tricritical, stephanov_fluc_qcd_crit, fluc_collective, signatures, charge_fluct, charge_fluct2, fluctuations_review_heiselberg, fluct3, fluct4, fluct5, methods_fluctuations, stephanov_thermal_fluc_pion, hijing_jet_study, gavin_pt_fluc,
ceres_pt,wa98_fluc, na49_fluc,
star_pt_fluc,
star_deta_dphi_cf_200, star_deta_dphi_cf, star_pt_fluc_excitation,
star_charge_fluc, 
star_balance, phenix_net_charge_fluc, phenix_pt_fluc,phenix_pt_2004,balance_theory,balance_distortions_jeon,
balance_distortions,balance_blastwave}.
Various theories predict that the production of a QGP phase in relativistic heavy-ion collisions could produce significant event-by-event correlations and fluctuations in temperature, transverse momentum, multiplicity, and conserved quantities such as net charge.

One such observable, the balance function, may be sensitive to the correlation of balancing charges \cite{balance_theory}.  For instance, for every particle of momentum $p$, there must be an anti-particle of momentum $p'$ with the opposite charge.
By means of a like-sign subtraction, the balance function can produce the distribution of relative momentum, $q = p - p'$, between the balancing charges.
Balance functions are sensitive to the mechanisms of charge formation and the subsequent relative diffusion of the balancing charges \cite{balance_theory}.  Balance functions are also affected by the freeze-out temperature and radial flow \cite{balance_distortions_jeon}. Remarkably, balance functions for central collisions have been shown to be consistent with blast-wave models where the balancing charges are required to come from regions with identical collective flow \cite{balance_blastwave}. The inferred high degree of correlation in coordinate space has been postulated as a signal for delayed hadronization \cite{balance_theory}, which would not allow charges the opportunity to separate in coordinate space.  The idea is that in central collisions a deconfined system of quarks and gluon is created, which cools and expands.  The observed balancing charges are then created when the deconfined system hadronizes, which reduces the effects of expansion and diffusion on the correlation of the balancing charges.
The same arguments were used in discussing charge fluctuations \cite{charge_fluct}.   Additionally, the same correlations would ensue if the charges were created early (on the order of 1 fm/$c$), but due to very limited diffusion, remained correlated at breakup.  Thus a narrowing of the balance function in central collisions implies delayed hadronization.  We have previously presented results for balance functions from Au+Au collisions at $\sqrt{s_{NN}}$ = 130 GeV for all charged particles and for identified charged pions \cite{star_balance}.  We observed that the balance function narrows in central Au+Au collisions for all charged particles and for identified charged pions.

UrQMD (Ultra-relativistic Quantum Molecular Dynamics, version 2.3) \cite{UrQMD} is an example of a model where charges are created early and there is significant diffusion in the subsequent evolution of the system.  Indeed, balance functions in terms of relative pseudorapidity or relative rapidity predicted by UrQMD do not exhibit narrowing in central collisions (see Section V).  Other models have been applied to predict balance functions. One model is based on a blast-wave and includes a thermal model with resonance decay \cite{Florkowski}.  This model cannot explain the narrowing of the balance function in central Au+Au collisions at $\sqrt{s_{NN}}$ = 130 GeV.  Another model attributes the narrow balance functions observed for central Au+Au collisions at $\sqrt{s_{NN}}$ = 130 GeV to quark-antiquark coalescence \cite{Bialas}.

Recently, the system size and centrality dependence of the balance function for all charged particles has been studied at $\sqrt{s_{NN}}$ = 17.3 GeV for p+p, C+C, Si+Si, and Pb+Pb collisions \cite{NA49_balance}.  The balance function for all charged particles narrows in central Pb+Pb collisions at 17.3 GeV and the widths of the balance functions for p+p, C+C, Si+Si, and Pb+Pb collisions scale with the number of participating nucleons.  The rapidity dependence and incident energy dependence of the balance function for all charged particles have been studied for Pb+Pb collisions from $\sqrt{s_{NN}}$ = 6.3 GeV to  $\sqrt{s_{NN}}$ = 17.3 GeV in Ref. \cite{NA49_balance_2007}.  The balance function is observed to narrow in central collisions for midrapidity, but does not narrow at forward rapidity.  The authors of Ref. \cite{NA49_balance_2007} show that the narrowing of the balance function in terms of $\Delta \eta$ in central collisions can be explained with the AMPT (A MultiPhase Transport) model incorporating delayed hadronization, while models such as HIJING and UrQMD fail to reproduce the observed narrowing.  We have recently presented a study of the longitudinal scaling of the balance function in Au+Au collisions at $\sqrt{s_{NN}}$ = 200 GeV \cite{longitudinal_balance}.

In this paper, we present new results for the balance function from p+p, d+Au, and Au+Au collisions at $\sqrt{s_{NN}}$ = 200 GeV.  These results have significantly better statistical accuracy than our previous measurements for Au+Au collisions at $\sqrt{s_{NN}}$ = 130 GeV and define the system size dependence of the balance function at $\sqrt{s_{NN}}$ = 200 GeV.  We present balance functions for all charged particles, charged pions, and charged kaons.  We also show the balance function in terms of several different variables that each have different sensitivities to different physical effects.   We compare our results with current theoretical predictions.

The balance function is calculated as:
\begin{eqnarray}
\label{BF}
B = \frac{1}{2} \left\{ \frac{\Delta_{+-} - \Delta_{++}}{N_{+}}+\frac{\Delta_{-+} - \Delta_{--}}{N_{-}}
\right\}
\end{eqnarray}
where $\Delta_{+-}$ in the case of identified charged pion pairs denotes the density (number divided by bin width) of identified charged pion pairs in a given range, e.g. relative rapidity $\Delta y = | y(\pi^{+}) - y(\pi^{-}) |$, and similarly for $\Delta_{++}$, $\Delta_{--}$, and $\Delta_{-+}$.   The terms $\Delta_{+-}$, $\Delta_{++}$, $\Delta_{--}$, and $\Delta_{-+}$ are calculated using pairs from a given event and the resulting distributions are summed over all events.  Specifically, $\Delta_{+-}$ is calculated by taking in turn each positive pion in an event and incrementing a histogram of $\Delta y$  with respect to all the negative pions in that event.  $\Delta_{+-}$ is then summed over all events.  A similar procedure is followed for $\Delta_{++}$, $\Delta_{--}$, and $\Delta_{-+}$. Eq. \ref{BF} is then used to calculate $B$ where 
$N_{+(-)}$ is the number of positive(negative) pions integrated over all events. The balance function is calculated for all events in a given centrality bin.  In the case of non-identified charged particle pairs, relative pseudorapidity ($\Delta\eta$) is used.  Balance functions using other variables are presented including the relative azimuthal angle, $\Delta \phi$, and the Lorentz invariant momentum difference between the two particles, $q_{\rm inv}$.  Balance functions in terms of $\Delta \phi$ are sensitive to flow and jet effects (See Section \ref{BDeltaPhi}).  Balance functions in terms of $q_{\rm inv}$ are sensitive more directly to the temperature of the emitting system (See Section \ref{Bqinv}).  In addition, balance functions are presented in terms of the components of $q_{\rm inv}$ in the rest frame of the particle pair: $q_{\rm long}$, in the beam direction; $q_{\rm out}$, in the direction of the transverse momentum of the particle pair; and $q_{\rm side}$, the direction perpendicular to $q_{\rm long}$ and $q_{\rm out}$.  Note that $q_{\rm inv}^2 = q_{\rm long}^2 + q_{\rm out}^2 + q_{\rm side}^2$ .

The width of the balance function is quantified in several ways.  For balance functions in terms of $\Delta \eta$, $\Delta y$, and $\Delta \phi$, the widths are calculated in terms of a weighted average.  For example the width of $B(\Delta \eta)$ is calculated as
\begin{eqnarray}
\label{WA}
\left\langle {\Delta \eta } \right\rangle  = \frac{{\sum\limits_{i = i_{{\rm{lower}}} }^{i_{{\rm{upper}}} } {B\left( {\Delta \eta _i } \right)\Delta \eta _i } }}{{\sum\limits_{i = i_{{\rm{lower}}} }^{i_{{\rm{upper}}} } {B\left( {\Delta \eta _i } \right)} }}
\end{eqnarray}
where $B\left( {\Delta \eta _i } \right)$ is the value of the balance function for the relative pseudorapidity bin $\Delta \eta _i$ and the sums are carried out from a beginning relative pseudorapidity bin $i_{{\rm{lower}}}$ to an ending bin $i_{{\rm{upper}}}$.  The lower bin is chosen to minimize contributions from background and final state interactions and the upper bin is the highest bin in $\Delta \eta$.  For balance functions in terms of $q_{\rm inv}$, we extract the width by fitting to a thermal distribution over a range in $q_{\rm inv}$.  Widths extracted from the measured balance functions are presented in Section V.

The data used in this analysis were measured using the Solenoidal Tracker at RHIC (STAR) \cite{TPCref,FEE}.  The Au+Au data were acquired during Run 7 at RHIC.  The p+p data were taken during Run 2 and the d+Au data were taken during Run 3.
The main detector was the Time Projection Chamber (TPC) located in a solenoidal magnetic field. The magnetic field magnitude was 0.50 T.
Events were selected according to the distance of their event vertex from the center of STAR.  Events were accepted within 1 cm of the center of STAR in the plane perpendicular to the beam direction.  Events were accepted with vertices within 10 cm of the center of STAR in the beam direction for Au+Au and within 15 cm for p+p and d+Au collisions.

Minimum-bias data were used in all cases.  Minimum-bias triggers for the Au+Au collisions were defined by the coincidence of two Zero Degree Calorimeters (ZDCs) \cite{ZDC} located $\pm$ 18 m from the center of the interaction region, along with an online cut on the Vertex Position Detectors (VPDs) that restricted accepted events to within 5 cm of the center of STAR in the beam direction.   For the Au+Au data set, 28 million events were analyzed.
For p+p and d+Au collisions, the trigger consisted of the two ZDCs combined with the Central Trigger Barrel (CTB) \cite{CTB}.  One million events were analyzed for the p+p data set and for the d+Au data set.
For Au+Au collisions, centrality bins were determined using the measured charged hadron multiplicity within the pseudorapidity range $|\eta| < 0.5$ as measured in the TPC.  The centrality bins were calculated as a fraction of this multiplicity distribution starting with the highest multiplicities.   The ranges used were 0-5\% (most central), 5-10\%, 10-20\%, 20-30\%, 30-40\%, 40-50\%, 50-60\%, 60-70\%, and 70-80\% (most peripheral).
For d+Au, three centrality bins were used, 0-20\%, 20-60\%, and 60-100\% determined by the multiplicity of charged particles originating from the primary collision vertex in the Forward Time Projection Chamber (FTPC), in the direction of the deuteron beam \cite{dAuSTAR}. Note that the pseudorapidity distribution for d+Au is not symmetric around $\eta$ = 0.  Each centrality was associated with a number of participating nucleons, $N_{\rm part}$, using a Glauber Monte Carlo calculation \cite{dAuSTAR,npart_ref}.  For p+p collisions, all multiplicities were used.

All tracks were required to have a distance of closest approach (DCA) to the measured event vertex of less than 3 cm.
Only charged particle tracks having more than 15 measured space points along the trajectory were accepted. The
ratio of the numbers of reconstructed space points to possible space points along the
track was required to be greater than 0.52. 
Charged pions and charged kaons were identified using the specific energy loss, $dE/dx$, along the track and the momentum, $p$, of the track.   Particle identification was accomplished by selecting particles whose specific energy losses were within two standard deviations of the energy-loss predictions for a given particle type and momentum.  Particle identification for pions (kaons) also included a condition that the specific energy loss should be more than two standard deviations away from the loss predicted for a kaon (pion).  In addition, electrons were excluded from the analysis for all cases by requiring that the specific energy loss for each track was more than one standard deviation away from the energy-loss predictions for electrons.

We estimated the systematic errors by comparing the results from Run 4 at RHIC with the results from Run 7 at RHIC, in which new tracking software was implemented.  We assign a 5\% systematic error on the extracted widths for the balance functions in terms of $\Delta \eta$ and $\Delta y$ and a 10\% systematic error on the extracted widths for the balance functions in terms of $q_{\rm inv}$ and $\Delta \phi$.

In this paper, we present an overview of the acceptance and efficiency of STAR in Section \ref{Acceptance} because the balance functions we present here are not corrected for acceptance and efficiency.  This section includes detailed track cut specifications.  We then present the balance functions for all the measured systems in Section \ref{BalanceFunctions}.  We compare some of the results with blast-wave model \cite{balance_blastwave} and HIJING (version 1.38) \cite{hijing} predictions in Section \ref{ComparisonModels}.  We then extract the widths of the balance functions and examine the systematics of these widths in Section \ref{Widths}.  Our conclusions are presented in Section \ref{Conclusions}.

\section{Data Acceptance and Efficiency}
\label{Acceptance}

\begin{figure}
\includegraphics[width=3.0in]{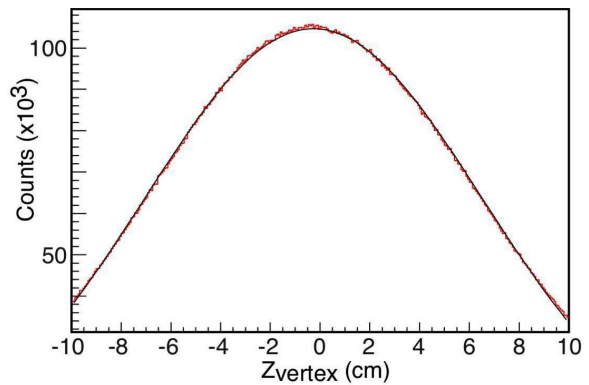}
\caption{\label{fig:nfig01}(Color online)  The distribution of the reconstructed position of the event vertex along the beam direction for events from Au+Au collisions at $\sqrt{s_{NN}}$ = 200 GeV.  The solid curve is a Gaussian fit with a mean of -0.27 cm and a standard deviation of 6.81 cm.}
\end{figure}

Here we outline the major acceptance and efficiency corrections necessary to compare any model calculation with the balance function results presented in this paper.  The pseudorapidity cut for all cases is $|\eta| < 1.0$.  The position of the vertex for each event along the beam direction affects the pseudorapidity acceptance of STAR.  The distribution of event vertices along the beam direction is shown in Fig.~\ref{fig:nfig01}.  The solid curve in Fig.~\ref{fig:nfig01} corresponds to a Gaussian fit with a mean of -0.27 cm and a standard deviation of 6.81 cm.  The distributions of event vertices in the beam direction for p+p and d+Au have a standard deviation of approximately 25 cm. 

\begin{figure}
\includegraphics[width=3.0in]{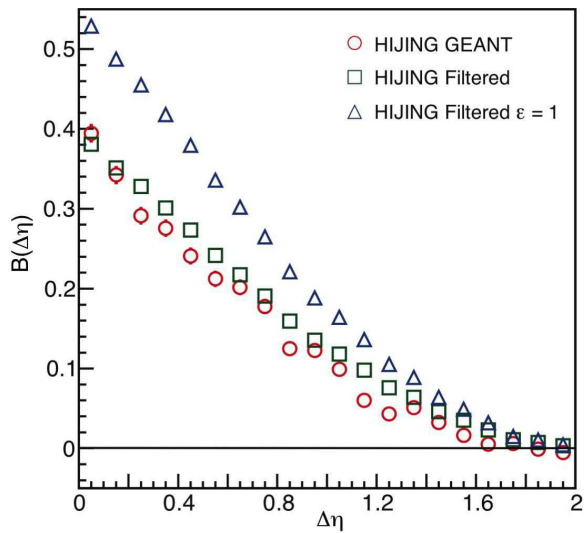}
\caption{\label{fig:nfig02}(Color online)  Calculated balance functions for all charged particles from central Au+Au collisions at $\sqrt{s_{NN}}$ =  200 GeV using HIJING.  The open circles depict HIJING events passed through GEANT and event reconstruction.  The open squares show HIJING events filtered with the acceptance and efficiency cuts described in the text.  The open triangles show HIJING events filtered with the acceptance cuts only.  When not shown, the statistical errors are smaller than the symbol size.}
\end{figure}

For the balance functions for all charged particles, we used a $p_{\rm t}$ cut of $0.2 < p_{\rm t} < 2.0$ GeV/$c$.  For identified particles, we used a $p_{\rm t}$ cut of $0.2 < p_{\rm t} < 0.6$ GeV/$c$.  For the high $p_{\rm t}$ measurements for $B(\Delta \phi)$, we used a $p_{\rm t}$ cut of $1.0 < p_{\rm t} < 10.0$ GeV/$c$.  The DCA cut of 3 cm partially suppressed particles resulting from weak decays.   The probability of accepting a charged particle in the fiducial volume of the TPC (including particle decay) is 90\% for charged particles with $p_{\rm t} > 0.2$ GeV/$c$.  
The efficiency for reconstructing a charged pion in our acceptance varies from 80\% in central collisions to 95\% in peripheral collisions. More details can be found in Refs. \cite{dAuSTAR}, \cite{Calderon}, and \cite{STAR_identified_spectra}.  We also suppressed electrons, resulting in the removal ($<$ 5\%) of pions in the momentum range $0.20 < p < 0.25$ GeV/$c$.  The electron cuts removed approximately 30\% of the identified kaons in the momentum range $0.4 < p < 0.8$ GeV/$c$.  To check these acceptance and efficiency corrections, we present balance functions based on 90k central HIJING events passing our event cuts that have been passed through GEANT and full event reconstruction.  We compare those results with our filtered HIJING calculations in Fig.~\ref{fig:nfig02}.  Filtered means that we apply our acceptance cuts in $\eta$ and $p_{\rm t}$ as well as the efficiency cuts listed above.  In addition, we present the filtered HIJING calculations with no efficiency correction ($\epsilon = 1$), but with all acceptance cuts applied. We see that the filtered HIJING results are similar to the full GEANT-filtered HIJING results within errors.  The widths of all three sets of HIJING data are the same within errors.

\section{Balance Functions}
\label{BalanceFunctions}

The balance functions $B(\Delta\eta)$ and $B(\Delta y)$ can be related to the correlation in rapidity of produced charge/anti-charge pairs.  By comparing PYTHIA calculations for p+p collisions with the results of a model describing a pion gas in which the opposite-charge pion pairs are assumed to be created together in space-time, the authors of Ref.~\cite{balance_blastwave} show that the balance functions from p+p collisions were wider than those from a thermal model.  Furthermore, they show that
the transport model RQMD (Relativistic Quantum Molecular Dynamics) \cite {RQMD}, in which the hadrons are created during the first 1 fm/$c$ after the collision, predicts that the balance function is wider in more central Au+Au collisions, which is the opposite of the experimental trend \cite{star_balance}.  We further observe that the transport model UrQMD \cite{UrQMD} predicts that the widths of the balance function in terms of $\Delta \eta$ and $\Delta y$ shows no centrality dependence for $\sqrt{s_{NN}}$ = 200 GeV Au+Au collisions (see Section V).

The authors of Ref.~\cite{balance_blastwave} make the point that the observed width of the balance function in terms of relative rapidity, $\sigma_{y}$, is a combination of the rapidity spread induced by thermal effects, $\sigma_{\rm therm}$, and the separation of the balancing partners of the charge/anti-charge pair in coordinate space.  The authors of Ref.~\cite{balance_theory} stated this relationship as $\sigma_{y}^{2} = \sigma_{\rm therm}^{2}+4\beta \ln{(\tau/\tau_{0})}$, where $\beta$ is a diffusion constant, $\tau$ is the proper time after the initial collision when the charge/anti-charge pair is created, and $\tau_{0}$ is a characteristic time on the order of 1 fm/$c$.  After the initial collision, the width of the balance function decreases because the thermal width narrows as a result of cooling, while diffusion tends to increase the width of the balance function.  If production of the charge/anti-charge pairs occurs at early times, then scattering and expansion affects the partners of the charge/anti-charge pair during the entire lifetime of the system. The diffusion term is then large and significantly broadens the observed balance function.  If the production of charge/anti-charge pairs occurs late, the time during which the partners of the charge/anti-charge pair are exposed to scattering and expansion is small, which makes the effect of diffusion negligible.  Thus, in the case of late production of the charge/anti-charge pairs, the width of the balance function is determined by the thermal width.
In Ref.~\cite{tonjes_PhD}, the dependence of these model calculations on delayed hadronization is demonstrated for a range of model assumptions.  The model calculations show that the longer hadronization is delayed, the narrower is the balance function.

In this section, we show the measured balance functions for p+p, d+Au, and Au+Au collisions at $\sqrt{s_{NN}}$ = 200 GeV.  We present balance functions for all charged particles, for charged pions, and for charged kaons.  
Throughout this paper, plotted balance functions based on Eq. \ref{BF} for data have been corrected by
subtracting the balance functions calculated using mixed events.
This subtraction corrects for differences between the acceptances for positive and negative particles.
Mixed events are created for each colliding system by grouping the events according to bins in centrality and bins in the position of the reconstructed vertex of the event along the beam direction.  For the Au+Au data set, ten centrality bins and five vertex bins were used.  For the p+p data, five bins in event vertex position were used.  No mixed events were created for the d+Au results because we only present results for $B(\Delta\eta)$ for all charged particles, which did not require mixed event subtraction.

A set of mixed events is created by taking one track from an event, selected according to the bin in centrality and the bin in event vertex position.  A mixed event includes no more than one track from any observed event.  This new mixed-event data set has the same number of events with the same multiplicity distribution as the original data set but all correlations are removed.  $B(\Delta \eta)$ and $B(\Delta y)$ calculated from mixed events are always zero for all centralities and for all $\Delta \eta$ and $\Delta y$.  However, the balance functions in terms of $\Delta \phi$ calculated using mixed events are not always zero.  The difference between the the behavior of positively charged particles and negatively charged particles crossing the boundary between TPC sectors causes the balance functions in terms of $\Delta \phi$ calculated with mixed events to be non-zero.  This effect is most pronounced in central collisions where the particle density is the highest.  These variations of $B(\Delta \phi)$ correspond to multiples of the azimuthal separation of the sector boundaries of the TPC ($\Delta \phi = 2\pi/12$ = 0.52).
Residual effects can still be seen in balance functions in terms of $\Delta \phi$ in the most central bins even after mixed event subtraction at $\Delta \phi$ values corresponding to multiples of the azimuthal separation of the TPC sector boundaries. 

For most of the measured systems, we also present balance functions calculated from shuffled events.  These shuffled events are produced by randomly shuffling the charges of the particles in each event.  The shuffled events thus have all the momentum correlations and the total charge observed in the original event, but the charge-momentum correlations are removed.  Because shuffling uniformly distributes a particle's balancing partner across the measured phase space,
$B(\Delta \eta)$ and $B(\Delta y)$ calculated using shuffled events can be used to gauge the widest balance functions that one can measure using the STAR acceptance for the system under consideration.  Balance functions calculated with shuffled events have the same integral as the balance functions calculated with the original events.  One exception for the shuffled events relates to balance functions calculated using low multiplicity events, specifically the results for $B(\Delta y)$ and $B(q_{\rm inv})$ for charged kaon pairs.  The balance functions calculated by shuffling low multiplicity events are not significantly different from the original events, because exchanging the positive and negative balancing partners has no effect on the resulting balance function.  Therefore, in the case of low multiplicity events, we create the shuffled events by sampling the parent distributions for the variable in question.  The resulting shuffled balance function using sampling has an integral equal to one.  The shuffled balance functions using sampling are scaled by the integral of the original balance function.  We verified that the shuffled events created using the sampling technique agree with the shuffled data in the case of high multiplicity events, specifically for $B(\Delta y)$ and $B(q_{\rm inv})$ for charged pion pairs.

\subsection{Balance Functions in Terms of $\Delta \eta$ and $\Delta y$}

\subsubsection{Au+Au at $\sqrt{s_{NN}}$ = 200 GeV}

\begin{figure}
\includegraphics[width=3.25in]{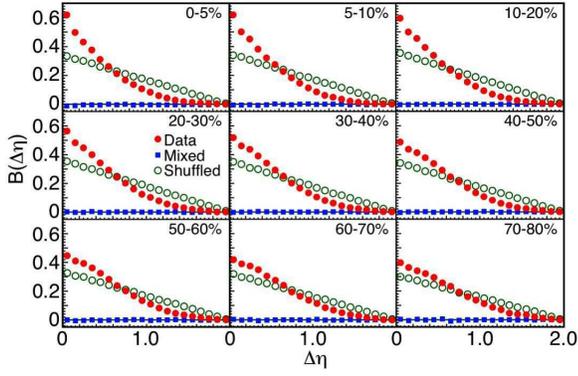}
\caption{\label{fig:nfig03}(Color online) The balance function in terms of $\Delta\eta$ for all charged particle pairs from Au+Au collisions at $\sqrt{s_{NN}}$ = 200 GeV for nine centrality bins.}
\end{figure}

Fig.~\ref{fig:nfig03} shows the balance function in terms of $\Delta \eta$ for all charged particles from Au+Au collisions at $\sqrt{s_{NN}}$ = 200 GeV for nine centrality bins.   The balance function gets narrower as the collisions become more central. The balance function for mixed events is zero for all centralities and $\Delta \eta$.  The balance function for shuffled events is significantly wider than the measured balance functions.  Model predictions show that inter-pair correlations (e.g. HBT and final state interactions) should be significant for $\Delta \eta < 0.1$ \cite{balance_distortions}. 

\begin{figure}
\includegraphics[width=3.25in]{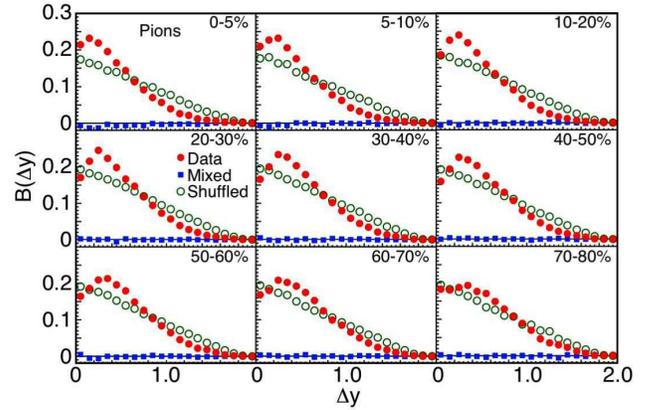}
\caption{\label{fig:nfig04}(Color online)  The balance function in terms of $\Delta y$ for identified charged pion pairs from Au+Au collisions at $\sqrt{s_{NN}}$ = 200 GeV for nine centrality bins.}
\end{figure}

\begin{figure}
\includegraphics[width=3.25in]{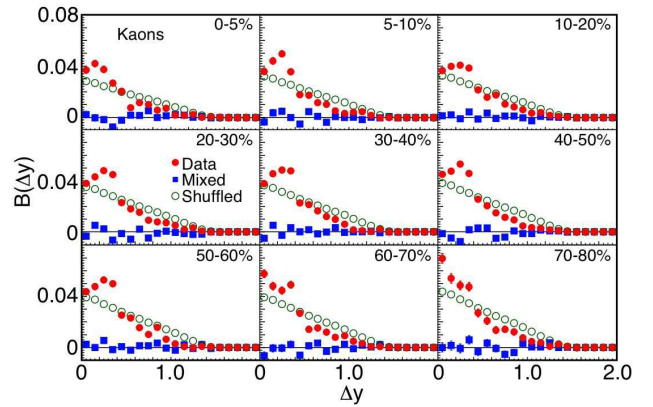}
\caption{\label{fig:nfig05}(Color online)  The balance function in terms of $\Delta y$ for identified charged kaon pairs from Au+Au collisions at $\sqrt{s_{NN}}$ = 200 GeV for nine centrality bins.}
\end{figure}

Figs.~\ref{fig:nfig04} and \ref{fig:nfig05} show the balance functions for identified charged pion pairs and kaons pairs, respectively, for Au+Au collisions at $\sqrt{s_{NN}}$ = 200 GeV for nine centrality bins as a function of the relative rapidity.  The balance function for identified pion pairs gets narrower in central collisions.  The lower magnitude of the balance function for pion pairs and kaon pairs compared with the balance function for all charged particles is due to the fact that the efficiency of observing an identified pion or a kaon is lower than for unidentified charged particles. The balance function calculated from mixed events is zero for all centralities and $\Delta y$ for both pions and kaons.  The balance functions calculated using shuffled events are substantially wider than the measured balance functions.  The discontinuity in $B(\Delta y)$ for kaons around $\Delta y$ = 0.4 visible at all centralities is due to $\phi$ decay, which was verified using HIJING calculations.
Model predictions show that inter-pair correlations should be significant for $\Delta y < 0.2$ \cite{balance_distortions}.  These effects scale with the multiplicity and thus are more apparent in central collisions.

\subsubsection{p+p and d+Au at $\sqrt{s_{NN}}$ = 200 GeV}

\begin{figure}
\includegraphics[width=2in]{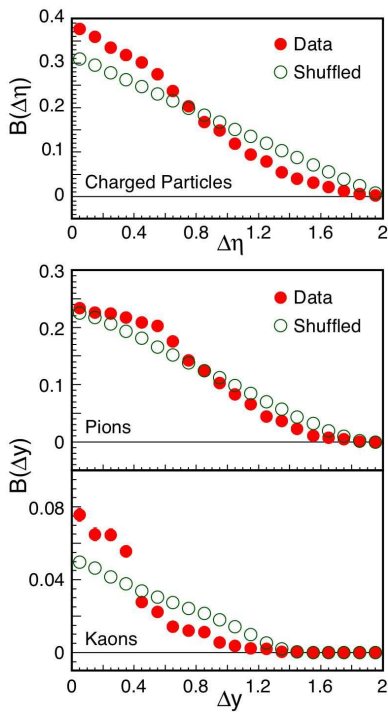}
\caption{\label{fig:nfig06}(Color online) The balance function for p+p collisions at $\sqrt{s}$ = 200 GeV.  The top panel shows the balance function for all charged particles in terms of $\Delta \eta$.  The bottom panel gives the balance function for charged pion pairs and charged kaon pairs in terms of $\Delta y$.}
\end{figure}

To investigate the system-size dependence of the balance function and to provide a nucleon-nucleon reference for the balance functions extracted from Au+Au collisions, we measured the balance functions for p+p and d+Au collisions at $\sqrt{s_{NN}}$ = 200 GeV.  Fig.~\ref{fig:nfig06} shows the balance functions for all charged particles for p+p collisions at  $\sqrt{s}$ = 200 GeV.  The balance functions for p+p collisions are integrated over all observed event multiplicities to allow comparison with centrality-selected d+Au and Au+Au results.  Note that the width of the balance function in terms of $\Delta \eta$ for p+p collisions is independent of the multiplicity of tracks in the event. The top panel of Fig.~\ref{fig:nfig06} shows the balance function for all charged particles in terms of $\Delta \eta$. In the bottom panel of Fig.~\ref{fig:nfig06}, the balance functions are shown for identified charged pion pairs and identified charged kaon pairs in terms of $\Delta y$ from p+p collisions at $\sqrt{s}$ = 200 GeV.   The balance function for mixed events is zero for all $\Delta \eta$ and all $\Delta y$. The observed shapes of the balance functions for the identified charged pions and kaons are similar to those observed in peripheral (70 - 80\%) Au+Au collisions.  The fact that the balance function for kaon pairs has a lower magnitude than the balance function for pion pairs reflects the lower efficiency for identifying charged kaons versus identifying charged pions in STAR.

\begin{figure}
\includegraphics[width=2in]{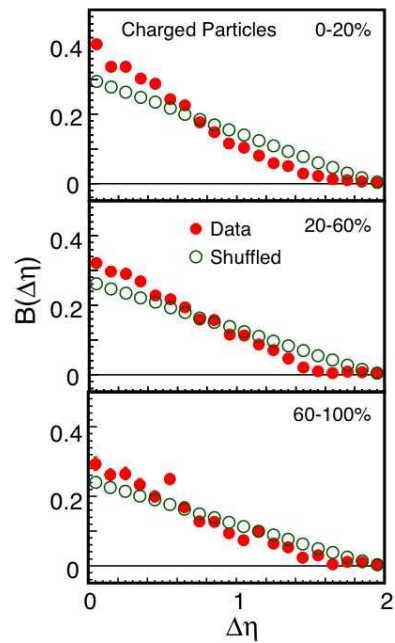}
\caption{\label{fig:nfig07}(Color online)  The balance function in terms of $\Delta\eta$ for all charged particles from d+Au collisions at $\sqrt{s_{NN}}$ = 200 GeV for three centrality bins.}
\end{figure}

Fig.~\ref{fig:nfig07} shows the balance functions in terms of $\Delta \eta$ for all charged particles from d+Au collisions at $\sqrt{s_{NN}}$ = 200 GeV for three centrality bins, 0-20\%, 20-60\%, and 60-100\%.

\subsection{Balance Functions in Terms of $q_{\rm inv}$}
\label{Bqinv}

The balance function in terms of $\Delta \eta$ and $\Delta y$ is
observed to narrow in central collisions and model calculations have
been used to interpret this narrowing in terms of delayed hadronization
\cite{balance_theory,balance_distortions_jeon,balance_distortions,balance_blastwave}. 
However, in a thermal model, the width of the balance function in terms
of $\Delta \eta$ and $\Delta y$ can be influenced by radial flow.  In the absence
of detector efficiency and acceptance considerations, the width of the balance
function in terms of the Lorentz invariant momentum difference between the
two particles,
$q_{\rm inv}$, is determined solely by the breakup temperature, if the
balancing charges are emitted from the same position in coordinate
space.   However, when detector acceptance is taken into account, some 
dependence on collective flow is introduced~\cite{balance_distortions}.
Thus, analyzing the balance function in terms of
$q_{\rm inv}$ avoids some of the complications associated with collective
flow, and the balance function calculated with a breakup temperature
should be the narrowest possible balance function if the particles are
assumed to be emitted from the same position in coordinate space. In
addition, contributions to the balance function from the decay of
particles are more identifiable.  For example, the decay of $K_S^0$
produces a sharp peak in $B(q_{\rm inv})$ for charged pions, while the
contribution to $B(\Delta y)$ for charged pions from the decay of 
$K_S^0$ is spread out over several bins in $\Delta y$.

To study balance functions in terms of $q_{\rm inv}$, we use identified
charged pions and identified charged kaons.  For pion pairs, we observe
a peak from the decay $K^{0}_{S} \rightarrow \pi^{+} + \pi^{-}$.  For
kaon pairs, we observe a peak from the decay $\phi \rightarrow K^{+} +
K^{-}$.  These peaks are superimposed on the balance function of
correlated charge/anti-charge pairs not resulting from the decay of a
particle.

\subsubsection{Au+Au at $\sqrt{s_{NN}}$ = 200 GeV}

\begin{figure}
\includegraphics[width=3.4in]{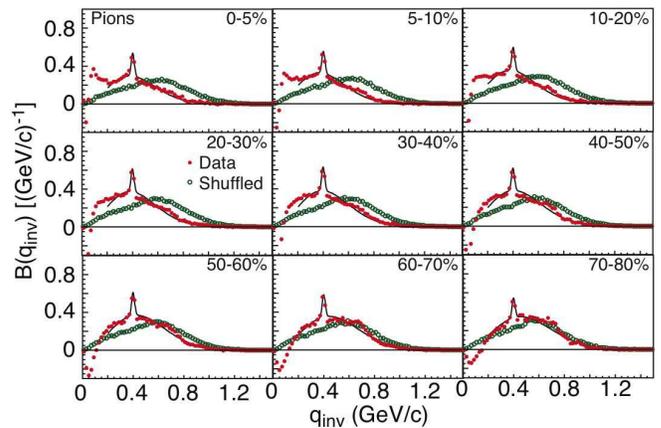}
\caption{\label{fig:nfig08}(Color online) The balance function
in terms of $q_{\rm inv}$ for charged pion pairs from Au+Au collisions at
$\sqrt{s_{NN}}$ = 200 GeV in nine centrality bins.  Curves correspond to a thermal distribution (Eq. \ref{thermal})
plus $K_{S}^{0}$ decay.}
\end{figure}

Fig.~\ref{fig:nfig08} shows the balance function for identified charged pions in terms of $q_{\rm inv}$ for Au+Au collisions at $\sqrt{s_{NN}}$ = 200 GeV for nine centrality bins. These balance functions have been corrected by subtracting the balance functions calculated using mixed events.  These mixed events are not zero for all $q_{\rm inv}$ because of differences in the tracking at TPC sector boundaries for opposite charges.  The balance functions calculated for mixed events integrate to zero as one would expect and the subtraction of the mixed events from the measured balance functions does not affect the integral of the resulting balance functions.   At each centrality, a peak is observed corresponding to charged pion pairs resulting from $K_{S}^{0} \to \pi ^{+} + \pi ^{-}$.  The solid curves represent a fit consisting of the sum of two terms.  The first term consists of a non-relativistic thermal distribution of the form 
\begin{eqnarray}
\label{thermal}
B(q_{\rm inv})=aq_{\rm inv}^2e^{-q_{\rm inv}^2/(2\sigma^{2})}
\end{eqnarray}
where $a$ is a constant, the pre-factor $q_{\rm inv}^2$ accounts for the phase-space effect, and $\sigma$ is a width parameter.  The second term of the fit is a Gaussian distribution in $q_{\rm inv}$ describing the $K_S^0$ decay.  Note that no peak from the decay of the $\rho^{0}$ is visible in central collisions around $q_{\rm inv}$ = 0.718 GeV/$c$ where one would expect to observe the $\rho^{0}$.  This non-observation of the $\rho^{0}$ is in contrast to HIJING, which predicts a large $\rho^{0}$ peak, as is demonstrated in Section \ref{ComparisonModels}.  The $\rho^{0}$ peak is visible in the most peripheral collisions, which is consistent with our previous study of $\rho^{0}$ production at higher $p_{\rm t}$ \cite{star_rho}.  The authors of Ref.~\cite{balance_blastwave} attribute the apparent disappearance of the $\rho^{0}$ in central collisions to the cooling of the system as it expands, which lowers the production rate of $\rho^{0}$ compared with pions.  The measured balance functions for pions are distinctly different from the balance functions calculated using shuffled events.  In particular, the sharp peak from the $K_S^0$ decay is not present in the balance functions calculated using shuffled events. 

\begin{figure}
\includegraphics[width=3.25in]{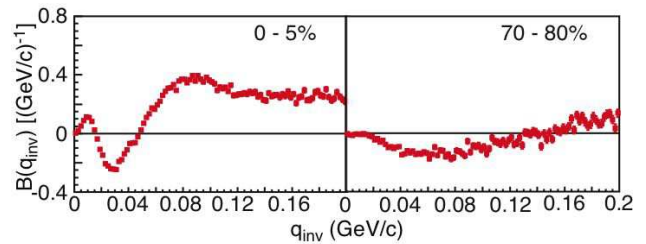}
\caption{\label{fig:nfig09}(Color online)  The balance function in terms of
$q_{\rm inv}$ for charged-pion pairs in two centrality bins over the range $0 < q_{\rm inv} < 0.2$ GeV/$c$.}
\end{figure}

HBT/Coulomb  effects are visible for $q_{\rm inv} < 0.2$ GeV/$c$ in
Fig.~\ref{fig:nfig08}.
Fig.~\ref{fig:nfig09} shows the balance function over the range of $0 < q_{\rm inv} < 0.2$ GeV/$c$ for the most central bin (0 - 5\%) and the most peripheral bin (70 -80\%).
The Coulomb force pulls opposite
charges closer together and pushes same charges apart, leading to an enhancement of opposite-sign and a
suppression of same-sign pairs at small $q_{\rm inv}$.  This effect leads to
a rise in the balance function at small $q_{\rm inv}$, which is larger in
central collisions, where the long-range Coulomb force affects more
particles \cite{balance_blastwave}.  In peripheral collisions, because the Coulomb interaction is
less important and the HBT correction is larger because of the smaller
source size, the Coulomb enhancement disappears and the balance function becomes
negative at small $q_{\rm inv}$ \cite{balance_blastwave}.

\begin{figure}
\includegraphics[width=3.4in]{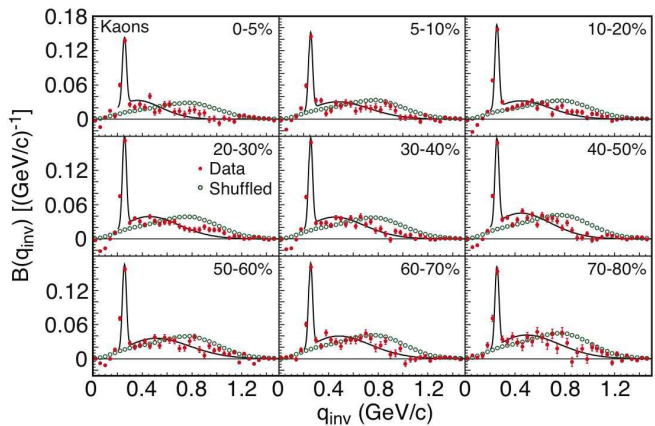}
\caption{\label{fig:nfig10}(Color online)  The balance function
in terms of $q_{\rm inv}$ for charged kaon pairs from Au+Au collisions at
$\sqrt{s_{NN}}$ = 200 GeV in nine centrality bins.  Curves correspond to a thermal (Eq. \ref{thermal})
distribution plus $\phi$ decay.}
\end{figure}

Fig.~\ref{fig:nfig10} shows the balance function for identified
charged kaons in terms of $q_{\rm inv}$ for Au+Au collisions at $\sqrt{s_{NN}}$ = 200 GeV in
nine centrality bins.  These balance functions were corrected by
subtracting mixed events as was done for the charged pion results.  At
each centrality, a peak is observed corresponding to charged kaon pairs
resulting from $\phi \to K ^{+} + K ^{-}$.  The solid curves represent
fits consisting of a non-relativistic thermal distribution (Eq. \ref{thermal})
plus a Gaussian distribution in $q_{\rm inv}$ for the $\phi$ decay. 
HBT/Coulomb effects at low $q_{\rm inv}$ for kaon pairs are not as strong as those observed
for pion pairs.  The measured balance functions are distinct from the balance functions calculated from shuffled
events.

Several differences between $B(q_{\rm inv})$ for charged pions and charged
kaons are evident.  The observed HBT/Coulomb effects at low $q_{\rm inv}$
are much stronger for pions than for kaons.  The HBT/Coulomb  effects
for pions change dramatically with centrality while the HBT  effects for
kaons are small and change little with centrality.   The overall
normalization for kaons is lower than the overall normalization for
pions, reflecting the lower efficiency for detecting identified kaons. 
The contribution to $B(q_{\rm inv})$ for pions from $K_S^0$ decay 
is approximately 7\%, independent of centrality. The contribution to 
$B(q_{\rm inv})$ for kaons from $\phi$ decay is approximately 50\%, 
independent of centrality.

\subsubsection{p+p at $\sqrt{s}$ = 200 GeV}

\begin{figure}
\includegraphics[width=3in]{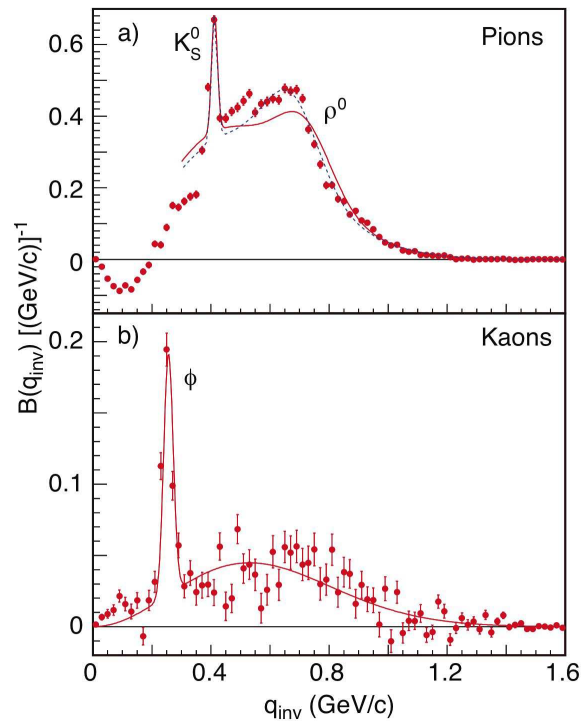}
\caption{\label{fig:nfig11}(Color online)  The balance function in terms of $q_{\rm inv}$ for charged pion pairs [part a)] and charged kaon pairs [part b)] from p+p collisions at $\sqrt{s}$ = 200 GeV integrated over all multiplicities. Solid curves correspond to a thermal distribution (Eq. \ref{thermal}) plus $K_{S}^{0}$ and $\rho^{0}$ decay for pions and $\phi$ decay for kaons.  The dashed curve for pions represents a fit to a thermal distribution (Eq. \ref{thermal}) plus $K_{S}^{0}$ decay and $\rho^{0}$ decay, with the $\rho^{0}$ mass shifted down by 0.04 GeV/$c^2$.}
\end{figure}

Fig.~\ref{fig:nfig11} shows the balance functions in terms of $q_{\rm inv}$ for p+p collisions at $\sqrt{s}$ = 200 GeV.   Fig.~\ref{fig:nfig11}a shows the balance function for charged pion pairs
and Fig.~\ref{fig:nfig11}b shows the balance function for charged kaon pairs. 
The solid curves are thermal fits (Eq. \ref{thermal}) plus a peak for $K_{S}^{0}$ 
and $\rho^{0}$ decay in the case of charged pions, and for $\phi$ decay 
in the case of charged kaons.  The thermal fit does not reproduce the
charged pion results, while it works well for the charged kaon data. The
mass of the $\rho^{0}$ used in the fit shown for pion pairs was assumed
to be 0.77 GeV/$c^2$.  A better fit can be obtained if the mass of the
$\rho^{0}$ is lowered by 0.04 GeV/$c^2$, as was observed previously in
studies of $\rho^{0}$ production in p+p collisions at $\sqrt{s}$ = 200
GeV \cite{star_rho}.  This fit is shown as a dashed curve in the upper
panel of Fig.~\ref{fig:nfig11}.  Note that the $\rho^{0}$ peak visible in
$B(q_{\rm inv})$ for pions from p+p collisions is not observed in
$B(q_{\rm inv})$ for pions from central Au+Au collisions, but is observed for pions from peripheral Au+Au collisions, as shown in
Fig.~\ref{fig:nfig08}.

\subsection{Balance Function in Terms of Components of $q_{\rm inv}$}

Here we present results for the three
components of $q_{\rm inv}$. These components are $q_{\rm long}$, the component
along the beam direction; $q_{\rm out}$, the component in the direction of
the transverse momentum of the observed pair; and $q_{\rm side}$, the component perpendicular to
$q_{\rm long}$ and $q_{\rm out}$.

Analysis of the balance function for these three components can address
the question of what causes the balance function to narrow in
central Au+Au collisions.  In a thermal model where the balancing
particles are emitted from the same position in coordinate space, the
widths would be identical for the three components.  On the other hand,
charge separation associated with string dynamics should result in
balance functions that are wider in $q_{\rm long}$ than in $q_{\rm side}$ or
$q_{\rm out}$ \cite{balance_distortions,balance_blastwave}.   Also because the velocity gradient is much higher in the
longitudinal direction, diffusion should broaden the balance function
more in $q_{\rm long}$ \cite{balance_blastwave}.

\begin{figure}
\includegraphics[width=3.25in]{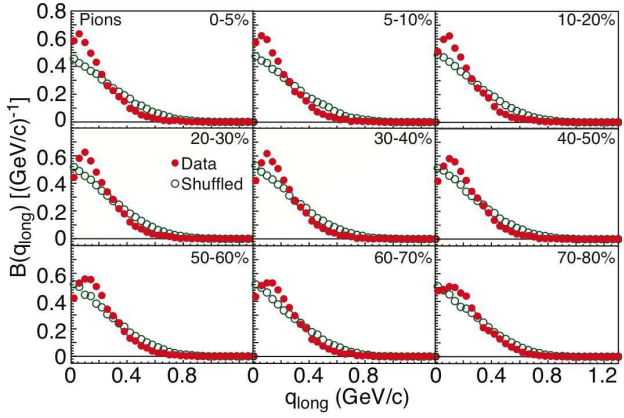}
\caption{\label{fig:nfig12}(Color online)  The balance function
in terms of $q_{\rm long}$ for charged pion pairs from Au+Au collisions at
$\sqrt{s_{NN}}$ = 200 GeV in nine centrality bins.}
\end{figure}

\begin{figure}
\includegraphics[width=3.25in]{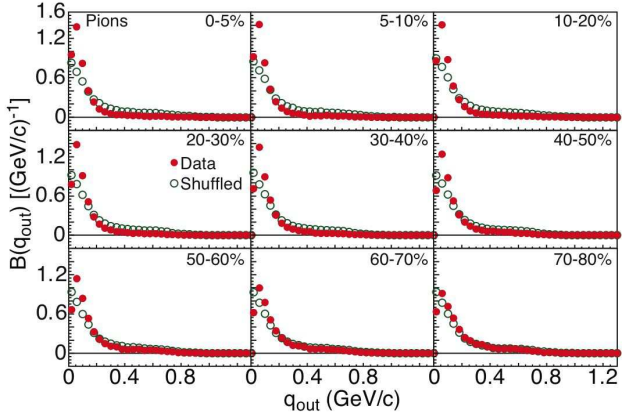}
\caption{\label{fig:nfig13}(Color online)  The balance function in terms of $q_{\rm out}$ for
charged pion pairs from Au+Au collisions at $\sqrt{s_{NN}}$ = 200 GeV
in nine centrality bins.}
\end{figure}

\begin{figure}
\includegraphics[width=3.25in]{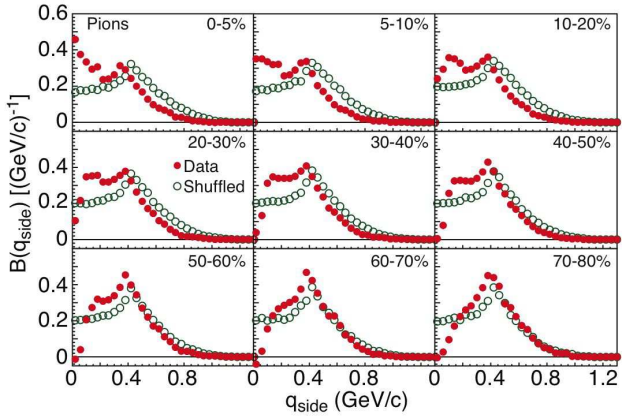}
\caption{\label{fig:nfig14}(Color online) The balance function in terms of $q_{\rm side}$ for
charged pion pairs from Au+Au collisions at $\sqrt{s_{NN}}$ = 200 GeV
for nine centrality bins.}
\end{figure}

Figs.~\ref{fig:nfig12}, \ref{fig:nfig13}, and \ref{fig:nfig14} show the balance functions for charged pion pairs
from Au+Au collisions at $\sqrt{s_{NN}}$ = 200 GeV in terms of $q_{\rm long}$,   $q_{\rm out}$, and  $q_{\rm side}$ respectively.
The balance functions calculated using mixed events
are subtracted from the measured balance functions.
The balance functions for all three components are narrower in central collisions than in peripheral collisions.

The balance
functions in terms of $q_{\rm side}$ do not look like those measured using $q_{\rm long}$
or $q_{\rm out}$ because the lower momentum cut-off of STAR strongly affects
$B(q_{\rm side})$ for $q_{\rm side} < 0.38$ GeV/$c$, which underscores the
importance of performing comparisons with models that have been put
through detailed efficiency and acceptance filters.

\subsection{Balance Functions in Terms of  $\Delta \phi$}
\label{BDeltaPhi}

The balance function in terms of $\Delta \phi$ may yield information related to transverse flow at freeze-out \cite{Bozek} and may be sensitive to jet production.
One might expect that jet-like phenomena would involve the emission of
correlated charge/anti-charge pairs at small relative azimuthal angles.
We present balance functions for all charged particles with $0.2 <
p_{\rm t} < 2.0$ GeV/$c$ from Au+Au collisions at $\sqrt{s_{NN}}$ = 200 GeV as a function of the
relative azimuthal angle, $\Delta \phi$.  In addition, we present
$B(\Delta\phi)$ for all charged particles with $1.0 < p_{\rm t} < 10.0$
GeV/$c$ to enhance any possible jet-like contributions to the balance
function.

\begin{figure}
\includegraphics[width=3.25in]{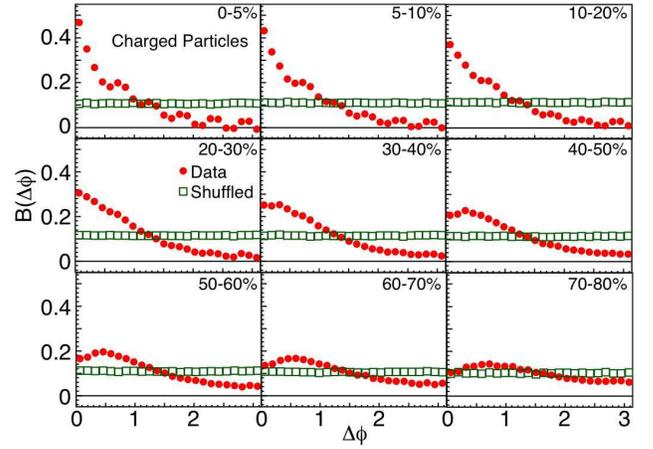}
\caption{\label{fig:nfig15}(Color online)  The balance function in terms of $\Delta \phi$ for
all charged particles with $0.2 < p_{\rm t} < 2.0$ GeV/$c$ from Au+Au
collisions at $\sqrt{s_{NN}}$ = 200 GeV in nine centrality bins. The
closed circles represent the real data minus the mixed events.}
\end{figure}

Fig.~\ref{fig:nfig15} shows the balance functions as a function of
$\Delta \phi$ for all charged particles with $0.2 < p_{\rm t} < 2.0$ GeV/$c$
in nine centrality bins.  The balance functions for mixed events were
subtracted.  Note that some structure in $\Delta \phi$ related to the 
sector boundaries of the STAR TPC is still visible
after the subtraction of the mixed events. We observe a peaking at $\Delta
\phi$ = 0 in central collisions, while in peripheral collisions, the
balance functions are almost flat.  Fig.~\ref{fig:nfig15} also shows
the balance functions calculated using shuffled events.  The balance
functions from shuffled events are constant with $\Delta\phi$ and show
no centrality dependence.

\begin{figure}
\includegraphics[width=3.25in]{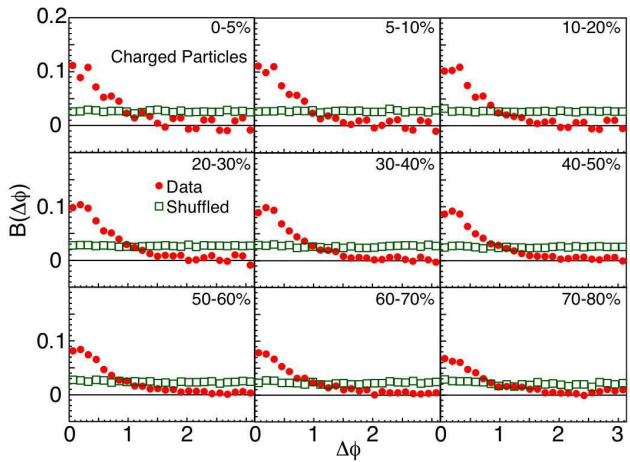}
\caption{\label{fig:nfig16}(Color online)  The balance function in terms of $\Delta \phi$ for
all charged particles with $1.0 < p_{\rm t} < 10.0$ GeV/$c$ from Au+Au
collisions at $\sqrt{s_{NN}}$ = 200 GeV in nine centrality bins. The
closed circles represent the real data minus the mixed events.}
\end{figure}

To augment this result, Fig.~\ref{fig:nfig16} presents balance functions 
in which we use only particles with $1.0 < p_{\rm t} < 10.0$ GeV/$c$. 
For this case, we see that the measured balance functions vary little
with centrality.  Again the balance functions calculated with shuffled
events are constant with $\Delta\phi$ and show no centrality dependence.
HIJING calculations for $B(\Delta\phi)$ for all charged particles with $0.2 < p_{\rm t} < 2.0$
GeV/$c$ exhibit little dependence on $\Delta\phi$, while HIJING
calculations for particles with $1.0 < p_{\rm t} < 10.0$ GeV/$c$ are peaked
at $\Delta\phi$ = 0, suggesting that the balance functions for this
higher $p_{\rm t}$ range show jet-like characteristics.

The dramatically tight correlations in $\Delta \phi$ in central
collisions of Au+Au shown in Fig.~\ref{fig:nfig15} are qualitatively
consistent with the radial flow of a perfect liquid.  In a liquid with very
short mean free path, the balancing particles would remain in close
proximity throughout the reaction.  A large mean free path, which would
necessitate a large viscosity, would damp the correlations in $\Delta
\phi$ \cite{Teaney}.  This trend is also consistent with a picture where charges are
not created until after the flow has been established.

\section{Comparison with Models}
\label{ComparisonModels}

\begin{figure}
\includegraphics[width=3in]{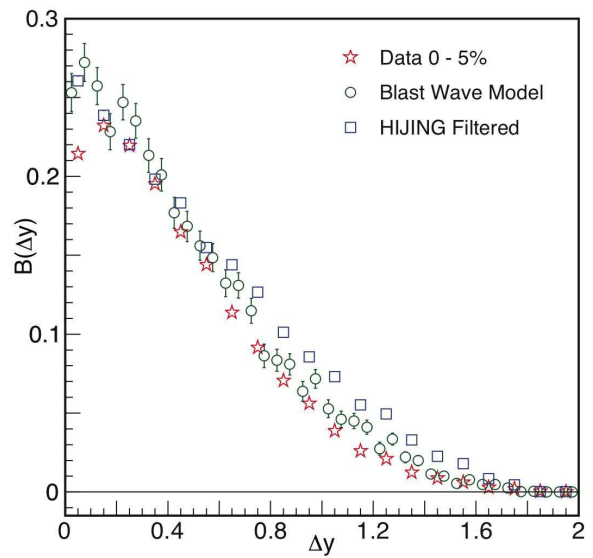}
\caption{\label{fig:nfig17}(Color online)  The balance function in terms of $\Delta y$ for charged pions from central collisions of Au+Au at $\sqrt{s_{NN}}$ = 200 GeV compared with predictions from the blast-wave model from Ref.~\cite{balance_blastwave} and filtered HIJING calculations taking into account acceptance and efficiency.}
\end{figure}

Fig.~\ref{fig:nfig17} compares the measured balance function $B(\Delta y)$ for charged pion pairs from central collisions of Au+Au at $\sqrt{s_{NN}}$ = 200 GeV to the predictions of the blast-wave model \cite{balance_blastwave} and to filtered HIJING calculations taking into account acceptance and efficiency.  The blast-wave model includes radial flow, emission of charge/anti-charge pairs of particles close together in space and time, resonances, HBT and Coulomb effects, strong force effects, inter-domain interactions, and a STAR experimental filter.  The blast-wave calculations shown in Fig.~\ref{fig:nfig17} include the acceptance cuts in the current paper.  The resulting absolute predictions of the blast-wave model agree well with the measured balance function.  In contrast, the balance function predicted by HIJING is significantly wider than the measured balance function.  The widths of the balance functions predicted by the blast-wave and HIJING are compared with the experimental values in Fig.~\ref{fig:nfig20}.

The width of the balance function predicted by the blast-wave model is close to the width observed in central collisions.  
The blast-wave model assumes that the charge/anti-charge pairs of particles are created close together in space and at the same time, and contains no scattering or longitudinal expansion that would widen the balance function in terms of $\Delta y$.  Thus, the agreement of the predicted width from the blast-wave model and the data is consistent with the idea of delayed hadronization in that delayed hadronization in central collisions would minimize the contribution of diffusion effects to the width of the balance function.

\begin{figure}
\includegraphics[width=3in]{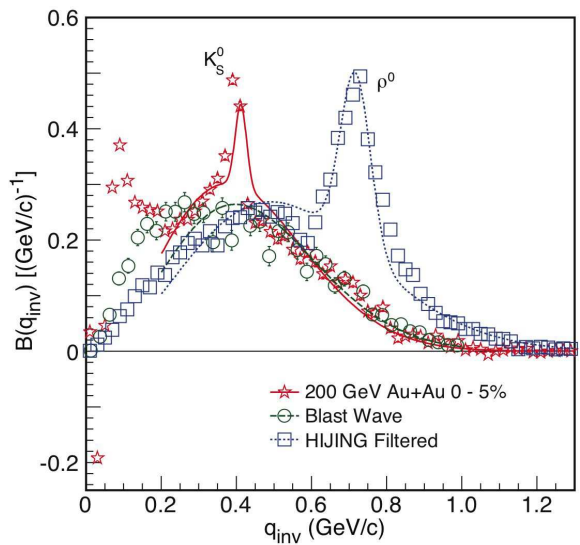}
\caption{\label{fig:nfig18}(Color online)  The balance function in terms of $q_{\rm inv}$ for charged pions from central collisions of Au+Au at $\sqrt{s_{NN}}$ = 200 GeV compared with predictions from the blast-wave model from Ref.~\cite{balance_blastwave} and predictions from filtered HIJING calculations including acceptance and efficiency. For the blast-wave calculations, HBT is not included and the decays of the $K_S^0$ and $\rho^{0}$ are not shown. }
\end{figure}

The balance function in terms of $q_{\rm inv}$ provides the most direct way to study the dependence of the balance function on temperature.  Fig.~\ref{fig:nfig18} compares the balance function in terms of $q_{\rm inv}$ for charged pion pairs from central collisions of Au+Au at $\sqrt{s_{NN}}$ = 200 GeV to the predictions of the blast-wave model and to filtered HIJING calculations.  For the blast-wave model calculations, HBT is not included and the decays of the $K^{0}$ and $\rho^{0}$ are not shown.  The solid curve for the data represents a fit comprised of a thermal distribution (Eq. \ref{thermal}) plus $K^{0}$ decay.  The dashed curve for the blast-wave model calculations represents a thermal fit (Eq. \ref{thermal}).  The dotted curve for the HIJING calculations represents a thermal distribution (Eq. \ref{thermal}) plus $\rho^{0}$ decay.  All the fits are carried out over a range in $q_{\rm inv}$ that is not affected by HBT/Coulomb effects.  The width extracted from the thermal fit to the blast-wave model calculations is compared with the width extracted from experimental data in Fig.~\ref{fig:nfig21}.  The blast-wave model reproduces the observed width in central collisions.  The HIJING calculations show a strong $\rho^{0}$ peak that is not present in the data.

Future analyses should be able to disentangle the effects of cooling and diffusion in driving the narrowing of the balance function.  Diffusive effects should largely manifest themselves in the $q_{\rm long}$ variable because the initial velocity is in the longitudinal direction and some creation mechanisms, such as strings, preferentially separate the pairs in the longitudinal direction.

\section{Balance Function Widths}
\label{Widths}

The balance functions presented in the previous section provide insight into the correlation of charge/anti-charge pairs in collisions at RHIC.  This approach complements the approach of studying these phenomena using charge-dependent correlation functions in two dimensions,
$(\Delta \eta,\Delta \phi)$ \cite{star_deta_dphi_cf_200, star_deta_dphi_cf}.
The balance function can be related to these correlation functions and to other two-particle observables.   $B(\Delta y)$ can be interpreted as the distribution of relative rapidities of correlated charge/anti-charge pairs.  The width of  $B(\Delta y)$ then can be used to determine whether correlated charge/anti-charge pairs of particles are emitted close together or far apart in rapidity.  The width of the balance function $B(q_{\rm inv})$ can be used to study thermal distributions because this balance function can be related to the temperature, and is largely unaffected by any radial expansion.

\begin{figure}
\includegraphics[width=3in]{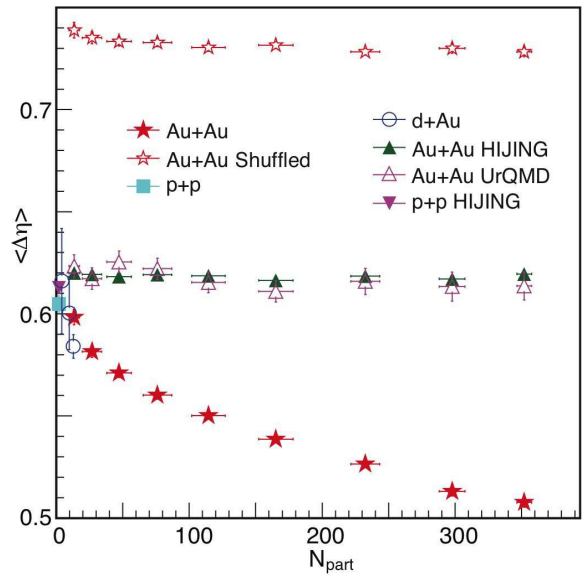}
\caption{\label{fig:nfig19}(Color online)  The balance function width $\langle \Delta \eta \rangle$ for all charged particles from Au+Au collisions at $\sqrt{s_{NN}}$ = 200 GeV compared with the widths of balance functions calculated using shuffled events.  Also shown are the balance function widths for p+p and d+Au collisions at $\sqrt{s_{NN}}$ = 200 GeV.  Filtered HIJING calculations are also shown for the widths of the balance function from p+p and Au+Au collisions. Filtered UrQMD calculations are shown for the widths of the balance function from Au+Au collisions.}
\end{figure}

To quantify the evolution of the balance functions $B(\Delta y)$ and $B(\Delta\eta)$ with centrality, we extract the width, $\langle \Delta y \rangle$ and $\langle \Delta \eta \rangle$, using a weighted average (Eq. \ref{WA}).  For $B(\Delta \eta)$, the weighted average is calculated for $ 0.1 \le \Delta \eta \le 2.0$ and for $B(\Delta y)$, the weighted average is calculated for $ 0.2 \le \Delta y \le 2.0$.

Fig.~\ref{fig:nfig19} shows the balance function widths for all charged particles from Au+Au, d+Au, and p+p collisions at $\sqrt{s_{NN}}$ = 200 GeV plotted in terms of the number of participating nucleons, $N_{\rm part}$.  In addition, we present the widths of the balance functions from Au+Au collisions for shuffled events.  The widths of the shuffled events are considerably larger than those from the measured data and represent the largest width we can measure using the STAR acceptance for the system under consideration. 

The balance function widths scale smoothly from p+p through the three centrality bins for d+Au and down to the nine Au+Au collision centrality data points. This figure also shows filtered HIJING calculations for p+p and Au+Au calculations for HIJING and UrQMD. The HIJING calculations for p+p reproduce the measured width. The Au+Au HIJING and UrQMD calculations, however, show little centrality dependence and are comparable to those calculated from the HIJING p+p simulations. This is despite the fact that HIJING does not predict any appreciable radial flow while UrQMD predicts radial flow in Au+Au collisions but less than that observed experimentally. This radial flow should produce a narrower balance function in central collisions where radial flow is the largest, while hadronic scattering should lead to a wider balance function.  The fact that the measured widths from Au+Au collisions narrow in central collisions is consistent with trends predicted by models incorporating late hadronization \cite{balance_theory,balance_blastwave}.

\begin{figure}
\includegraphics[width=3.25in]{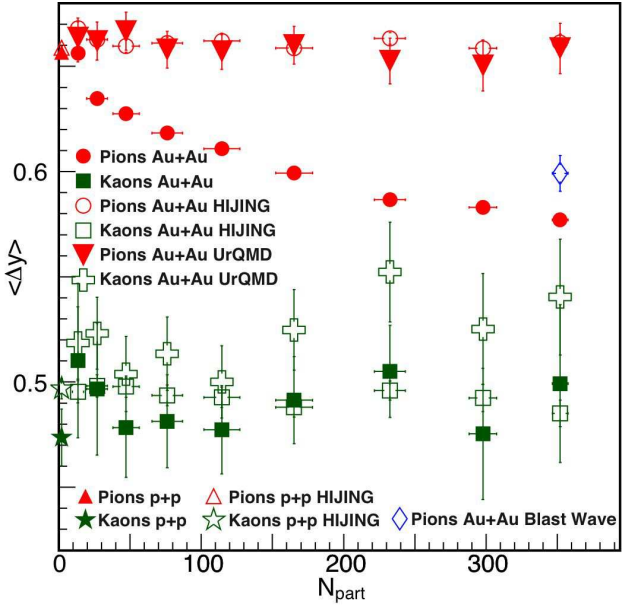}
\caption{\label{fig:nfig20}(Color online)  The balance function widths for identified
charged pions and charged kaons from Au+Au collisions at $\sqrt{s_{NN}}$
= 200 GeV and p+p collisions at $\sqrt{s}$ = 200 GeV.  Filtered HIJING
calculations are shown for the same systems.  Filtered UrQMD calculations are shown for Au+Au. Also shown is the width of the balance function for pions predicted by the blast-wave model of Ref. \cite{balance_blastwave}.}
\end{figure}

\begin{figure*}
\includegraphics[width=5.5in]{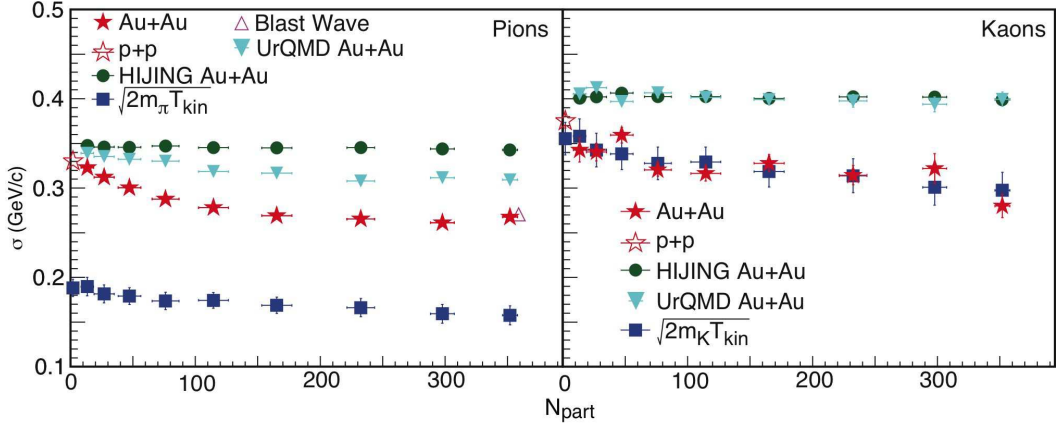}
\caption{\label{fig:nfig21}(Color online)  The balance function width $\sigma$ extracted from
$B(q_{\rm inv})$ for identified charged pions and kaons from Au+Au collisions at $\sqrt{s_{NN}}$ = 200 GeV 
and p+p collisions at $\sqrt{s}$ = 200 GeV using a thermal fit (Eq. \ref{thermal}) where $\sigma$ is the width.  Filtered HIJING and UrQMD calculations are shown for pions and kaons from Au+Au collisions at $\sqrt{s_{NN}}$ = 200 GeV.  Values are shown for $\sqrt{2mT_{\rm kin}}$ from Au+Au collisions, where $m$ is the mass of a pion or a kaon, and $T_{\rm kin}$ is calculated from identified particle spectra \cite{STAR_identified_spectra}.  The width predicted by the blast-wave model of Ref. \cite{balance_blastwave} is also shown for pions.}
\end{figure*}

Fig.~\ref{fig:nfig20} presents the widths of the balance function, $B(\Delta y)$, for identified charged pions and identified charged kaons from p+p collisions at $\sqrt{s}$ = 200 GeV and Au+Au collisions at $\sqrt{s_{NN}}$ = 200 GeV.  Also shown are filtered HIJING and UrQMD calculations.  For charged pions, the measured balance function widths for Au+Au collisions get smaller in central collisions, while the filtered HIJING and UrQMD calculations for Au+Au again show no centrality dependence.  The HIJING calculations for p+p collisions reproduce the observed widths.

In contrast, the widths of the measured balance function for charged kaons from Au+Au collisions show little centrality dependence.  The extracted widths for charged kaons are consistent with the predictions from filtered HIJING calculations and are consistent with the p+p results.  The widths for charged kaons predicted by UrQMD are somewhat larger than the data.  The agreement with HIJING and the lack of centrality dependence may indicate that kaons are produced mainly at the beginning of the collision rather than during a later hadronization stage \cite{balance_theory}.  The larger widths predicted by UrQMD for kaons may reflect the hadronic scattering incorporated in UrQMD, although the statistical errors are large for both the data and the model predictions.
 
Fig.~\ref{fig:nfig21} shows the widths extracted from $B(q_{\rm inv})$ for identified charged pions and kaons from Au+Au collisions at $\sqrt{s_{NN}}$ = 200 GeV and p+p collisions at  $\sqrt{s}$ = 200 GeV using a thermal distribution (Eq. \ref{thermal}) where $\sigma$ is the width.  The widths for the pions are somewhat smaller than the widths for the kaons, although the kaon widths have a large statistical error.  This width is related to the temperature of the system when the pions and kaons are formed.  Filtered HIJING calculations show no centrality dependence and predict a difference between the widths for pions and kaons.  The widths predicted by UrQMD for pions are smaller than those predicted by HIJING but are still larger than the measured widths.  In addition, the widths predicted by UrQMD for pions seem to show a centrality dependence, although it is not as strong as that for the data.  The widths predicted by UrQMD for kaons show no centrality dependence and agree with HIJING.

For a thermal system in the non-relativistic limit ($m \gg T$), the balance function has the functional form  given in Eq. \ref{thermal} where $\sigma=\sqrt{2mT}$.
For kinetic freeze-out temperatures $T \sim 0.1$ GeV \cite{STAR_identified_spectra}, kaons are non-relativistic, and this functional form was seen to describe the balance function in Fig.~\ref{fig:nfig10}.  Indeed, as seen in the right panel of Fig.~\ref{fig:nfig21}, the evolution in the width of the balance function may be understood in terms of the evolution of the freeze-out temperature as a function of centrality  \cite{STAR_identified_spectra}.

In the ultra-relativistic case ($m \ll T$), the balance function from a thermal system is exponential rather than Gaussian, $B(q_{\rm inv}) \sim q_{\rm inv}^2e^{-q_{\rm inv}/T}$.
The proper functional form for pions, being neither non-relativistic nor ultra-relativistic, is more complicated.  Indeed, we found that neither the Gaussian form nor the exponential form fully describe the pion balance
function in Fig.~\ref{fig:nfig08}.  Thus, to get a feeling for whether the evolution in freeze-out temperature can explain the narrowing of the balance function for pions, we turn to numerical calculations. Calculations in Ref. \cite{balance_distortions} show a 27\% reduction in the Gaussian width of  $B(q_{\rm inv})$ as the temperature is varied from 120 to 90 MeV, the temperatures inferred from fits to peripheral and central collisions, respectively \cite{STAR_identified_spectra}.  As seen in Fig.~\ref{fig:nfig21}, the measured width for peripheral (central) collisions is 0.33 GeV/$c$ (0.27 GeV/$c$), a 18\% reduction. Thus, the centrality evolution in freeze-out temperature may help explain much of the narrowing of the
balance function in terms of $q_{\rm inv}$ for pions as well as for kaons.  However, firm conclusions require more complete calculations including all detector effects.

\begin{figure}
\includegraphics[width=3.25in]{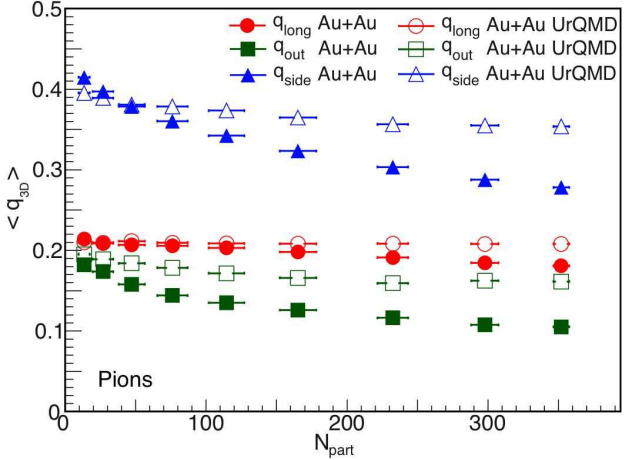}
\caption{\label{fig:nfig22}(Color online)
The widths for the balance functions for pions in terms of $q_{\rm long}$, $q_{\rm out}$, and $q_{\rm side}$ compared with UrQMD calculations.}
\end{figure}

Fig.~\ref{fig:nfig22} shows the widths of the balance functions in terms of $q_{\rm long}$, $q_{\rm out}$, and $q_{\rm side}$ for charged pion pairs in Au+Au collisions at $\sqrt{s_{NN}}$ = 200 GeV compared with the results of filtered UrQMD calculations.  These widths were extracted by taking the weighted average over the $q_{\rm long}$, $q_{\rm out}$, and $q_{\rm side}$ range from 0.0 to 1.3 GeV/$c$.  The width $\langle q_{\rm side} \rangle$ is larger than $\langle q_{\rm long} \rangle$ and $\langle q_{\rm out} \rangle$ because the lower $p_{\rm t}$ threshold of STAR affects it more strongly.  In the most peripheral collisions, the widths $\langle q_{\rm long} \rangle$ and $\langle q_{\rm out} \rangle$ are comparable to each other.  As the collisions become more central, both $\langle q_{\rm long} \rangle$ and $\langle q_{\rm out} \rangle$ decrease.  The change in $\langle q_{\rm long} \rangle$ is less than the change of $\langle q_{\rm out} \rangle$ with increasing centrality.  Thus it seems that the two transverse widths, $\langle q_{\rm out} \rangle$, and $\langle q_{\rm side} \rangle$, decrease in central collisions more strongly than the longitudinal width, $\langle q_{\rm long} \rangle$.  This may imply that string dynamics and diffusion due to longitudinal expansion may keep $\langle q_{\rm long} \rangle$ from decreasing as much in more central collisions \cite{balance_blastwave}.  The decrease in the transverse widths is consistent with the decrease in $T_{\rm kin}$ as the collisions become more central. In the most peripheral collisions, the widths predicted by UrQMD are consistent with the data.  As the collisions become more central, the predicted widths decrease slightly, but not as much as observed in the data.  This is consistent with results using the balance function in terms of $q_{\rm inv}$.  Additional theoretical input is required to draw more conclusions from the analysis of the balance function in terms of the components of $q_{\rm inv}$.

\begin{figure}
\includegraphics[width=3.25in]{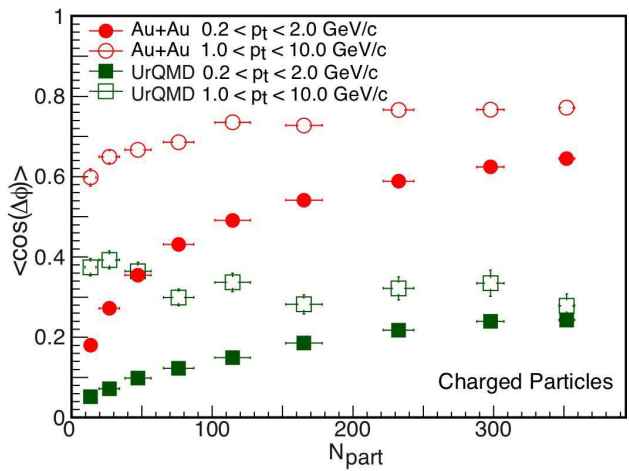}
\caption{\label{fig:nfig23}(Color online)  The weighted average cosine of the relative azimuthal angle, $\langle \cos{(\Delta \phi)} \rangle$,  extracted from $B(\Delta \phi)$ for all charged particles with $0.2 < p_{\rm t} < 2.0$ from Au+Au collisions at $\sqrt{s_{NN}}$ = 200 GeV and from all charged particles with $1.0 < p_{\rm t} < 10.0$ GeV/$c$, compared with predictions using filtered UrQMD calculations.}
\end{figure}

Fig.~\ref{fig:nfig23} shows the weighted average cosine of the relative azimuthal angle, $\langle \cos{(\Delta \phi)} \rangle$, extracted from the balance
functions $B(\Delta \phi)$ for all charged particles from Au+Au
collisions at $\sqrt{s_{NN}}$ = 200 GeV with $0.2 < p_{\rm t} < 2.0$ GeV/$c$ and $1.0 < p_{\rm t} < 10.0$ GeV/$c$. The values for $\langle \cos{(\Delta \phi)} \rangle$  are extracted over the range $0 \le \Delta \phi \le \pi$.  For the lower $p_{\rm t}$ particles, the balance function narrows dramatically in central collisions (large positive values of $\langle \cos{(\Delta \phi)} \rangle$).  The narrow balance functions observed in central collisions may be a signature of the flow of a perfect liquid, as discussed above.  For the higher $p_{\rm t}$ particles, $\langle \cos{(\Delta \phi)} \rangle$ in Au+Au collisions shows less centrality dependence.

Fig.~\ref{fig:nfig23} also shows UrQMD calculations for  $\langle \cos{(\Delta \phi)} \rangle$.  The predictions for the $0.2 < p_{\rm t} < 2.0$ GeV/$c$ data set are much lower than the measured values, which is consistent with the observation that UrQMD underpredicts radial flow.  The predictions for $\langle \cos{(\Delta \phi)} \rangle$ for the $1.0 < p_{\rm t} < 10.0$ GeV/$c$ data set show no centrality dependence and are also much lower than the measured values.

\section{Conclusions}
\label{Conclusions}

We have measured balance functions for p+p, d+Au, and Au+Au collisions at $\sqrt{s_{NN}}$ = 200 GeV for all charged particles, identified charged pions, and identified charged kaons.  We observe that the balance functions in terms of $\Delta \eta$ for all charged particles and in terms of $\Delta y$ and $q_{\rm inv}$ for charged pions narrow in central Au+Au collisions.  This centrality dependence is consistent with trends predicted by models incorporating delayed hadronization. The balance functions
$B(\Delta \eta)$ and $B(\Delta y)$ can be affected by radial flow while the balance function $B(q_{\rm inv})$ is largely unaffected by the implied reference frame transformation.  We observe that the system size dependence of the width of the balance function for charged particles scales with $N_{\rm part}$ as was observed at $\sqrt{s_{NN}}$ = 17.3 GeV \cite{NA49_balance}.  In contrast, HIJING and UrQMD model calculations for the width of the balance function in terms of $\Delta y$ or $\Delta \eta$ show no dependence on system size or centrality.

For charged kaons we observe that the width of the balance function $B(\Delta y)$ shows little dependence on centrality for Au+Au collisions at $\sqrt{s_{NN}}$ = 200 GeV.  This lack of dependence on centrality may indicate that strangeness is created early in the collision rather than in a later hadronization stage.  However, the fact that the balance function for kaons in terms of $q_{\rm inv}$ narrows in central collisions might be explained by the exclusion of the $\phi$ decay in the fits to $B(q_{\rm inv})$, while the $\phi$ decays are included in $B(\Delta y)$.

For both pions and kaons, the width of the balance function in $q_{\rm inv}$ decreases with increasing centrality.  This narrowing may be driven largely by the evolution of the kinetic freeze-out temperature with centrality.  This explanation is strengthened by the observation that the widths of the balance functions for pions in terms of the two transverse components of $q_{\rm inv}$, $q_{\rm out}$ and $q_{\rm side}$, decrease in central collisions.  However, more quantitative conclusions require more complete theoretical studies.

A comparison with a blast-wave model \cite{balance_blastwave} suggests that the balance function $B(\Delta y)$ for pion pairs in central Au+Au collisions at $\sqrt{s_{NN}}$ = 200 GeV is as narrow as one could expect, as the model assumed that the balancing charges were perfectly correlated in coordinate space at breakup.  This correlation might be explained either by having the charges created late in the reaction, thus denying them the opportunity to separate in coordinate space, or having them created early, but maintaining their close proximity through very limited diffusion. Whereas the first explanation is motivated by a picture of delayed hadronization, the idea of limited diffusion is consistent with the matter having a very small viscosity, which also requires a small mean free path. Furthermore, both these explanations account for the observation that the balance function narrows with centrality, since the breakup temperature, which determines the width, falls with increasing centrality.  The additional information provided here concerning the decomposition of the balance function into $q_{\rm out}$, $q_{\rm side}$, and $q_{\rm long}$ may provide the basis for a more stringent test of competing theoretical pictures.

We thank the RHIC Operations Group and RCF at BNL, the NERSC Center at 
LBNL and the Open Science Grid consortium for providing resources and 
support. This work was supported in part by the Offices of NP and HEP 
within the U.S. DOE Office of Science, the U.S. NSF, the Sloan Foundation, 
the DFG cluster of excellence `Origin and Structure of the Universe' of Germany, 
CNRS/IN2P3, STFC and EPSRC of the United Kingdom, FAPESP CNPq of Brazil, 
Ministry of Ed. and Sci. of the Russian Federation, NNSFC, CAS, MoST, and 
MoE of China, GA and MSMT of the Czech Republic, FOM and NWO of the 
Netherlands, DAE, DST, and CSIR of India, Polish Ministry of Sci. and 
Higher Ed., Korea Research Foundation, Ministry of Sci., Ed. and Sports of 
the Rep. Of Croatia, Russian Ministry of Sci. and Tech, and RosAtom of 
Russia.

\end{document}